\documentstyle[preprint,eqsecnum,aps,prd]{revtex}
\tighten
\begin{document}
\preprint{
Phys. Rev. D 57 (1998) 4812-4820
~~~~~~~~~~~~~~~~~~~~~~~~~~~~~~~~~~~~~~~~~~~~~~
gr-qc/9708024}
\draft

\title {
Monopole and electrically charged dust thin shells in general relativity:
Classical and quantum comparison of hollow and atomlike configurations
       }

\author{Konstantin G. Zloshchastiev\thanks{Electronic 
address: zlo@ff.dsu.dp.ua}
       }
\address{
Department of Theoretical Physics, Dnepropetrovsk University, 
Dnepropetrovsk 320625, Ukraine}

\date{Received: 12 August 1997 (LANL); 20 August 1997 (PRD)}
\maketitle

\def\bfl{\begin{flushleft}}
\def\efl{\end{flushleft}}
\def\bfr{\begin{flushright}}
\def\efr{\end{flushright}}
\def\bc{\begin{center}}
\def\ec{\end{center}}
\def\be{\begin{equation}}
\def\ee{\end{equation}}
\def\ba{\begin{eqnarray}}
\def\ea{\end{eqnarray}}
\def\nn{\nonumber }

\def\Kummer#1#2#3{\, \text{M}\left(#1,\,#2;\,#3 \right) } 
\def\KummerII#1#2#3{\, \text{U}\left(#1,\,#2;\,#3 \right) }
\def\BesselJ#1#2{\, \text{J}_{#1}\left(#2\right) } 
\def\BesselY#1#2{\, \text{Y}_{#1}\left(#2\right) }

\newcommand{\ben}{\begin{enumerate}}
\newcommand{\een}{\end{enumerate}}


\begin{abstract}
We give a comparative description of monopole and electrically 
charged spherically symmetric dust thin shells.
Herewith we consider two of the most interesting 
configurations: the hollow shell and shell, surrounding a body with 
opposite charge.
The classification of shells in accordance with the types of black holes and 
traversable wormholes is constructed.
The theorems for the parameters of turning points are proved.
Also for atomlike configurations the effects of screening 
(electrical case) and amplification (monopole case) of the internal mass  
by shell charge are studied.
Finally, one considers the quantum aspects; herewith, exact solutions 
of wave equations and bound states spectra are found.
\end{abstract}


\narrowtext

\section{INTRODUCTION}
\label{int}

Recently, investigations of the influence of cosmological phase transitions 
upon the
formation of topological defects, such as domain walls, strings, and 
monopoles, have been the focus of interest (see Ref.\ \cite{vil} for a 
review on 
these and related developments).
Monopoles forming as a result of gauge symmetry breaking have 
many properties of elementary particles; e.g., their energy is concentrated 
in a small region.

Barriola and Vilenkin \cite{bv} (see also Ref.\ \cite{dny} for additions) 
considered the
gravitational field of a global monopole formed as a result of 
global O(3) symmetry breaking.
The simplest model that gives rise to such monopoles is described by the 
Lagrangian
\be \label{eq1}
{\cal L} = \frac{1}{2}
           \left(
                 \frac{\partial \psi^a}{\partial x^b}
           \right)^2
          -\frac{\lambda}{4}
           \left(
                 \psi^a \psi^a - \frac{\eta^2}{8 \pi}
           \right)^2,
\ee
where $\psi^a$ is the isoscalar triplet, and $a=1,~2,~3$.
The topologically nontrivial self-supporting ansatz describing a monopole is
\[
\psi^a = \frac{\eta}{\sqrt{8 \pi}} \frac{x^a}{r}.
\]
The stress-energy tensor is given by
\be \label{eq2}
T^0_0=T^1_1= \frac{\eta^2}{8 \pi r^2},
\ee
where $r^2=x^a x^a$.
It was shown that spherically symmetric spacetime with the metric
(in the paper everywhere except for the sections devoted to 
quantum questions, we use the units $\gamma=c=1$, where $\gamma$ is the 
gravitational constant)
\be \label{eq3}
ds^2=-
     \left(
           1-\eta^2 -\frac{2 M}{r}
     \right) dt^2 +
     \left(
           1-\eta^2 -\frac{2 M}{r} 
     \right)^{-1} dr^2 +
     r^2 d\Omega^2
\ee
has the stress-energy tensor given by Eq.\ (\ref{eq2}) ($T^2_2=T^3_3=0$) 
and can approximately describe a global monopole 
with charge $\eta$ ($\eta^2 \ll 1$) and mass $M$ .
Notice that the total energy of this solution is divergent and the
presence of charge $\eta$
breaks the asymptotic flatness of the Schwarzschild field.
It makes the monopole an illusory object in the modern Universe
in the absence of cosmic phase transitions.
Nevertheless, if one considers the spherically symmetric gravitational 
collapse of the matter around the monopole, it can easily be seen that a 
black hole is formed.
Therefore, it would be very useful and instructive to study not only 
classical (for instance, the light deflection or perihelion-shift) or
semiclassical (Hawking radiation) effects, as was done in 
Ref.\ \cite{dny}, but also some features of the solution (\ref{eq3}) 
within the framework of the thin shell theory.

Thus, the aim of this paper is to study classical and quantum mechanical
properties of the thin shells with a global
monopole charge, and to compare them with the Reissner-Nordstr\"om ones.
Electrically charged hollow shells were considered by means of the 
Darmois-Israel formalism \cite{isr} in a lot of works \cite{ci,haw,ber}.
Nevertheless, we will pay attention both to the features of monopole charged 
configurations and to some insufficiently explored properties of 
electrically charged ones.

It is well known that the gravitational field produced by a static electric 
charge $e$ is described by the Reissner-Nordstr\"om metric
\be \label{eq4}
ds^2=-
     \left(
           1- \frac{2 M}{r} + \frac{e^2}{r^2}
     \right) dt^2 +
     \left(
           1- \frac{2 M}{r} + \frac{e^2}{r^2}
     \right)^{-1} dr^2 +
     r^2 d\Omega^2,
\ee
the stress-energy tensor for which is
\be \label{eq5}
T^0_0=T^1_1= -T^2_2=-T^3_3 =\frac{e^2}{8 \pi r^4}.
\ee

The key common element in Eqs.\ (\ref{eq2}) and (\ref{eq5}) 
is that the property $T^0_0=T^1_1$  allows one to obtain exact results
in the theory of spherically symmetric thin shells.
Thus we will consider the (2+1)-dimensional timelike dust layer $\Sigma$ 
with the surface energy-momentum tensor
\be  \label{eq6}
S_{ab}=\sigma u_a u_b,
\ee
where $\sigma$ is the surface mass density, $u_a$ are the components of 
tangent vector, and latin indexes denote surface tensors.

Let the metrics of the spacetimes outside $\Sigma^{\text{out}}$ and 
inside $\Sigma^{\text{in}}$ 
the shell be written in the form
\be     \label{eq7}
ds_{{\text{out}}\choose{\text{in}}}^2=-
   [1+\Phi^\pm (r)] dt_\pm^2 +
   [1+\Phi^\pm (r)]^{-1} dr^2 +
   r^2 d\Omega^2.
\ee
If one uses the shell proper time $\tau$, then the rest mass conservation 
law reads
\be  \label{eq8}
d \left(
         \sigma \sqrt{^{(3)}\!g} 
  \right) + \sqrt{^{(3)}\!g}
  \left(
        T_{\text{out}}^{\tau n} - T_{\text{in}}^{\tau n}
  \right) d\tau = 0,
\ee
where $T^{\tau n}=T^{\alpha\beta} u_\alpha n_\beta$ is the
projection of the four-dimensional stress-energy tensor on the tangent and 
normal vectors, $n_\alpha n^\alpha = -u_\alpha u^\alpha =1$, and $^{(3)}\!g$ is 
the absolute value of the determinant of the surface metric
\be  \label{eq9}
^{(3)}\!ds^2 = -d\tau^2 + R^2 d\Omega^2,
\ee
where $R=R(\tau)$ is the circumference radius of a shell.

The Lichnerowicz-Darmois-Israel junction conditions give us not only
the conservation law of the (effective) rest mass $m(R)$ of dust
but also the 
equations of motion of a shell:
\be                                                         \label{eq10}
\epsilon_+ \sqrt{\dot R^2 + 1+\Phi^+(R)} -
\epsilon_- \sqrt{\dot R^2 + 1+\Phi^-(R)} = -\frac{m(R)}{R},
\ee
\be                                                         \label{eq11}
m(R) = 4\pi\sigma(R) R^2,
\ee
where $\dot R=d R/d \tau$, 
$\epsilon_\pm = \text{sgn} \left(  \sqrt{\dot R^2 + 
1+\Phi^\pm(R)}  \right)$.
The choice of the signs $\epsilon_+$ and $\epsilon_-$ 
leads either to the impossibility of a junction or to a 
division of shells into 
the two classes of black hole (BH) and wormhole (WH) types \cite{ber,gb}.
It can be correctly determined for any fixed $\Phi^\pm(R)$ only.
                                        
Further, it can be shown that for spacetimes (\ref{eq7}) with the 
property $T^0_0=T^1_1$, the identity
\be      \label{eq12}
T^{\tau n} =0
\ee
is valid.
From  Eqs.\ (\ref{eq8}), (\ref{eq11}), and (\ref{eq12}) it 
follows that
\[
m=\text{const}.
\]

This paper is arranged as follows.
Section \ref{sec-esh} is devoted to electrically charged shells.
Two of the most interesting
configurations, namely, the hollow and atomlike shells, are studied. 
In Sec. \ref{sec-msh} analogous 
questions for monopole shells are considered.\footnote{Note that
we will not study the cosmological monopole bubbles 
($\sigma=\text{const}$; see \cite{kms} and references therein), the 
dynamics of which is determined by 
the junction conditions of the de Sitter and Barriola-Vilenkin  metrics.}
Also the quantum properties for the considered systems are given in 
these  sections.
Finally, some concluding remarks are made in Sec. \ref{sec-con}.

\section{ELECTRICALLY CHARGED SHELLS} \label{sec-esh}

For the maximum possible configuration  expressions (\ref{eq7}) 
and (\ref{eq10}) yield
\be                                     \label{eq13}
ds_\pm^2=-
     \left(
           1- \frac{2 M_\pm}{r} + \frac{e_\pm^2}{r^2}
     \right) dt_\pm^2 +
     \left(
           1- \frac{2 M_\pm}{r} + \frac{e_\pm^2}{r^2}
     \right)^{-1} dr^2 +
     r^2 d\Omega^2,                                          
\ee
\be                          \label{eq14}
\epsilon_+ \sqrt{\dot R^2 + 1 - \frac{2 M_+}{R} + 
\frac{e_+^2}{R^2}} -
  \epsilon_- \sqrt{\dot R^2 + 1 - \frac{2 M_-}{R} + 
\frac{e_-^2}{R^2}} =
  -\frac{m}{R}.                                              
\ee

In the general case they correspond to the shell surrounding a massive charged 
body with charge $e_-$ and mass $M_-$.
Then the external observable total mass (energy) and charge are $M_+$ and
$e_+$, respectively.
Below we consider the special cases.

\subsection{Hollow shell} 
\label{sec-esh-h}

This is a shell, inside of which the spacetime is flat, i.e., 
$M_-=0,~e_-=0$.
Supposing $M_+=M,~e_+=e$, we rewrite Eq.\ (\ref{eq14}) in the form
\be  \label{eq15}
\epsilon_+ \sqrt{\dot R^2 + 1 - \frac{2 M}{R} + 
\frac{e^2}{R^2}} - \epsilon_- \sqrt{\dot R^2 + 1} = -\frac{m}{R}.
\ee

Analysis of this expression under the restriction of the positive mass $m$ 
and radius $R$ gives us the classification of shells, represented in 
Table \ref{tab-1}, where $R_{\text{extr}}=e^2/2 M$ for this case.
From the table it can easily be seen that the radius value can have an
influence on the classification.
The radius $R_{\text{extr}}$  separates, in a natural way, the shells, which 
are pertinent to the worlds of microscopic and ordinary scales.
For instance, for a shell having the charge of $10^{20}$ electron's one 
and mass $10^{20}~\text{g}$, we obtain 
$R_{\text{extr}} \sim 10^{-37}~\text{cm}$.
For a shell with charge and mass of an electron 
$R_{\text{extr}} \sim 10^{-5}~\text{cm}$.
Hence it is clear that the case $R<R_{\text{extr}}$ describes the
microscopic shells, but the case $R>R_{\text{extr}}$ is connected with the shells 
of ordinary or astronomic sizes.
Note that in accordance with Table \ref{tab-1}, the shells of BH type have 
two subtypes depending on $\epsilon_\pm$.
Generally speaking, it is connected with the relativity of the concepts 
``outside'' and ``inside'' the shell.
If $\epsilon_+=\epsilon_-=1$ (in the table it corresponds to $R>R_{\text{extr}}$), 
then the shell occupies less than a half of the Universe 
(a similar situation was described in Ref.\ \cite{gb}).

For the second subtype $\epsilon_+=\epsilon_-=-1$ 
($R<R_{\text{extr}}$), and the shell occupies more than half of the 
Universe.
If one takes into account a small value of the radius $R_{\text{extr}}$, 
then the conclusion follows that a great part of the Universe is hidden 
under the shell of a microscopic radius (from the standpoint of an 
``external'' observer in $\Sigma^-$).
Such kinds of ideas have been exploited already for a long time, but in this 
case as will be shown below, these shells do not really exist at
the physical restriction $m>0$.

Now we formulate the statement, the proof of which is apparent from 
Eq.\ (\ref{eq15}).

{\it Proposition 1.}\ \
Let us have a hollow spherically symmetric dust shell with rest 
mass $m$ and charge $e$.
In addition, the condition of positivity of $M$ and $m$ is imposed.
Then
(i) the presence of charge increases the total mass $M$ of the shells of BH 
and WH types,
\be \label{eq16}
M=m \sqrt{1+\dot R^2} - \frac{m^2-e^2}{2 R};
\ee
(ii) the presence of charge increases the mass magnitude of black holes 
and wormholes with a throat radius equal to the horizon one, 
formed by such shells \cite{eq17proof},
\be \label{eq17}
M_H=\frac{m}{2}
    \left(
          1+\frac{e^2}{m^2}
    \right),
\ee
while the horizon radius is $R_H = M + \sqrt{M^2 - e^2}$.

In this connection it should be noted that the feature of shells reaching the 
external horizon $1 + \Phi^+(R_H)=0$ (so-called ``black'' shells) is that 
even if 
they have a nonzero radial velocity in a small neighborhood of the 
horizon point $R=R_H$, then the shell radius varies for an 
infinite time interval \cite{isr},
\[
d t_+ = \frac{\sqrt{ \dot R^2 +1 + \Phi^+(R)}} 
             {1 + \Phi^+(R)}  d \tau,
\]
from the viewpoint of a distant static observer in external 
spacetime $\Sigma^+$.
Thus, under an external observation the ``black'' shells are static.
If we assume also the turning point to be on the horizon, it will be a 
sufficient condition that the shell approach the horizon with 
respect to a comoving observer too. 

Below we formulate the turning point theorem, which can be very useful,
e.g., for a WKB quantization and analysis of the shell decay or nucleation
\cite{kms,aabs}.

{\it Theorem 1.}\ \ 
Let the conditions of the previous proposition be valid.
Then the shell has the turning point
\be \label{eq18}
R_s=\frac{1}{2} \frac{e^2-m^2}{M-m}
\ee
under the next restrictions of the total (mass) energy $M$\\
(a) if $\epsilon_+ = \epsilon_- = 1$, then $M \leq M_H$,\\
(b) if $\epsilon_+ = -\epsilon_- = -1$, then $M \geq M_H$,\\
where $M_H$ was defined by Eq.\ (\ref{eq17}).
Herewith, under the requirement of dominant charge $e^2 > m^2$ the
total energy hyperbolicity condition $M>m$ must be satisfied, and
under $e^2 < m^2$ the ellipticity condition $M<m$ must be satisfied.

{\it Proof:} 
The radius $R_s$ is determined by the single root of Eq.\ (\ref{eq16}) 
when $\dot R=0$:
\[
R_s=\frac{1}{2} \frac{e^2-m^2}{M- \epsilon_- m}.
\]
The conditions of its positivity and $\epsilon_- = 1$ \cite{eq17proof}
give the equality
\[
\text{sgn} (M-m) = \text{sgn} (m^2-e^2).
\]

It should be noted that in nonrelativistic theory we would obtain
the equality $\text{sgn} (M) = \text{sgn} (m^2-e^2)$.

Further, Eq.\ (\ref{eq16}) was obtained by a squaring of Eq.\ (\ref{eq15}),
and so we must again substitute $R_s$ in Eq.\ (\ref{eq15}) for
checking.
It gives us the additional constraint
\be                             \label{eq19}
\text{sgn} \left( 
                  e^2 - m^2 - 2 m (M-m)
            \right) = \epsilon_+,
\ee
from which it follows that
\ba
&&\epsilon_+ = 1,~~~M \leq \frac{e^2 + m^2}{2 m},     \label{eq20}\\
&&\epsilon_+ = -1,~~M \geq \frac{e^2 + m^2}{2 m}.     \label{eq21}
\ea
Comparing the RHS of the inequalities (\ref{eq20}) and (\ref{eq21}) with
Eq.\ (\ref{eq17}) we obtain the desired requirements (a) and (b). Q.E.D.

\subsection{Atomlike shell. Mass screening} 
\label{sec-esh-a}

Now we study the case when the shell surrounds a body of mass $M_-$ with 
an opposite charge.
Then the metrics of the spacetimes outside and inside the shell
will be described by expression (\ref{eq13}), in which we suppose
$e_-=e,~e_+=0$.
From the standpoint of an external observer in $\Sigma^+$ 
such a configuration will be 
electrically neutral, but an observer in $\Sigma^-$ will ``see'' 
the charge $e$.
According to Eq.\ (\ref{eq14}), the equation of motion can be written in the
form
\be  \label{eq22}
\epsilon_+ \sqrt{\dot R^2 + 1 - \frac{2 M_+}{R} } -
\epsilon_- \sqrt{\dot R^2 + 1 - \frac{2 M_-}{R} + 
\frac{e^2}{R^2}} =
-\frac{m}{R}.
\ee

We give the classification of these shells, similar to that done above.
One distinguishes the following two cases

(i) $M_+ < M_-$.
The classification is identical to that represented in 
the Table \ref{tab-1}, if 
one has to put $R_{\text{extr}}=e^2/2(M_- - M_+)$.

(ii) $M_+ \geq M_-$.
In this case, the classification does not depend on the radius 
(see Table \ref{tab-2}).

Further, it is easy to rewrite Eq.\ (\ref{eq22}) in the two forms
\be              \label{eq23}
M_+=M_- + \epsilon_- m\sqrt{
                              1-\frac{2 M_-}{R}+\frac{e^2}{R^2} +\dot R^2
                             }  -\frac{m^2+e^2}{2 R},          
\ee
\be             \label{eq24}
M_+=M_- + \epsilon_+ m\sqrt{
                              1-\frac{2 M_+}{R} +\dot R^2 } 
+ \frac{m^2-e^2}{2 R},                                          
\ee
which reflect the fact that the dynamics of a spherically symmetric thin 
shell can be reduced to the dynamics of a (1+1)-dimensional relativistic 
point particle in an external static field (see Refs. \cite{hb,aabs}).

Now we consider shells with the radius equal to the external horizon  
$R=2 M_+$ (see the comments after proposition 1).
From Eq.\ (\ref{eq22}) one obtains a square equation with respect to 
their mass,
\be \label{eq25}
M_+^2 -M_+ M_- - \frac{m^2-e^2}{4} =0,
\ee
having the two roots
\be \label{eq26}
2 M_+ = M_- \pm \sqrt{M_-^2 + m^2-e^2}.
\ee
Taking into account the positivity of $M_+$ and $M_-$, we choose the two 
possible variants
\be                   \label{eq27}
m^2>e^2~ \mapsto~  M_+ = \frac{1}{2}
                          \left(
                                M_- + \sqrt{M_-^2 + m^2-e^2}
                          \right),                             
\ee
\be                   \label{eq28}
m^2<e^2,~(m^2-e^2)^2 \leq M_-^4~
           \mapsto~  M_+ = \frac{1}{2}
                          \left(
                                M_- \pm \sqrt{M_-^2 + m^2- e^2}
                          \right).
\ee

Thus, the dependence of the observable mass on the internal mass is 
nonlinear and includes a charge.
It is easy to see that it leads to the appearance of some effects,
e.g., the phenomenon of the screening of internal mass by a shell charge.

\subsection{Quantum shells}
\label{sec-esh-q}

It is well known that several approaches to the quantization of spherically
symmetric thin shells in general relativity exist \cite{not}.
It is connected with the different ways of constructing the Hamilton formalism
and choice of gauge conditions \cite{hb}.
For instance, in Refs.\ \cite{ber,bkkt} the proper time gauge is used.
Another way \cite{haw,hkk} is to introduce a super-Hamiltonian on an
extended minisuperspace that leads to the Wheeler-DeWitt equation.
The last method has two main worthwhile qualities.
There are the evident agreement with the correspondence principle
(thereby we have physical vizualization) and an opportunity to 
obtain exact results .
It should be noted that the important feature of the shell theory is its 
time invariance.
Therefore, we can consider a shell as a stationary quantum system with a 
single radial degree of freedom, in contradistinction to a thick layer of 
matter, which has an infinite number of degrees of freedom.

Now we study electrically charged quantum shells in the most general case, 
when
spacetimes outside and inside the shell are given by Eq.\ (\ref{eq13}).
Here and in Sec. \ref{sec-msh-q} we use Planckian units.

If one has to put the momentum $\Pi=m\dot R$, then  it is
possible to rewrite Eq.\ (\ref{eq14}) as
the energy-momentum conservation law of a (1+1)-dimensional relativistic 
point particle:
\be         \label{eq29}
\left(
      \Delta M -\frac{e_+^2 -e_-^2}{2 R}
\right)^2 -\Pi^2 =
m^2
\left(
      1 - \frac{M_+ +M_-}{R} + \frac{2 (e_+^2 + e_-^2) - m^2}{4 R^2}
\right),
\ee
where $\Delta M=M_+ - M_-$.
We perform a direct quantization of this conservation law. 
Substituting the operator $\hat\Pi=-i\partial_R$ for the momentum $\Pi$, 
we consider eigenfunctions 
and eigenvalues of the self-adjointed operator given by the constraint 
(\ref{eq29}) 
(which corresponds to the singular Sturm-Liouville problem) and obtain the 
equation for the spatial wave function $\Psi (R)$  of a shell:
\be   \label{eq30}
\Psi^{\prime\prime} +
\left(
      \chi^2 +\frac{a}{R} +\frac{b}{R^2}
\right)\Psi =0,
\ee
where
\ba                                        
&&\chi=\sqrt{(\Delta M)^2-m^2},  \nn\\
&&a=m^2 (M_+ +M_-) - \Delta M (e_+^2 -e_-^2),            \label{eq31} \\
&&4 b= (e_+^2 -e_- ^2)^2 - 2 m^2(e_+^2 +e_-^2) + m^4.                \nn
\ea
Let us consider the asymptotics of this equation.\\ 
(a)$R \rightarrow \infty$.  
We have
\be                                \label{eq32}
\Psi^{\prime\prime} + \chi^2\Psi =0,                      
\ee
\be                                         \label{eq33}
\Psi = C_+ e^{i \chi R} + C_- e^{-i \chi R}. 
\ee
(b) $R \rightarrow 0$. 
\be                   \label{eq34}
R^2 \Psi^{\prime\prime} + b\Psi =0,                        
\ee
\be                    \label{eq35}
\Psi = c_+ R^{\lambda_{+1}} + c_- R^{\lambda_{-1}}, 
\ee
where $C_\pm,~c_\pm$ are integration constants,
\[
\lambda_\zeta =\frac{1+\zeta\alpha}{2},                      
\]
\be                     \label{eq36}
\alpha = \sqrt{1-m^4 +2 m^2(e_+^2 +e_-^2) - (e_+^2 -e_- ^2)^2}, 
\ee
and the sign $\zeta=\pm 1$ denotes the additional splitting of the solutions
(for more details and discussion see \cite{hkk}).
Equation (\ref{eq30}) can be written as the Whittaker's equation
\be                    \label{eq37}
\frac{d^2 \Psi}{d(2 i x)^2} +
\left(
      -\frac{1}{4} - \frac{i\beta}{2ix} +\frac{1-
      \alpha^2}{4(2ix)^2}
\right)\Psi = 0,
\ee
where
\ba
&&x=\chi R,                                                      
\nn\\
&&\beta =
\frac{
     m^2(M_+ + M_-) - \Delta M (e_+^2 -e_-^2)
     }{2\chi}.                                              
\label{eq38}
\ea
The solutions of this equation vanishing at zero are the functions
\be                               \label{eq39}
\Psi_\pm (\alpha,~\beta;~x>0) = e^{\pm i x} 
x^{\lambda_\zeta}
\Kummer{\lambda_\zeta + i\beta}{2\lambda_\zeta}{2 i x},
\ee
where $\Kummer{a}{b}{z}$ is the regular Kummer confluent hypergeometric 
function \cite{as}.

Now we derive the asymptotics of this solution at a large radius $R$ in the 
form \cite{haw}
\be                               \label{eq40}
\Psi_\pm (\alpha,~\beta;~x\rightarrow +\infty) = 
\Gamma(2\lambda_\zeta)
\left[
     \left(
       \frac{i}{2}
     \right)^{\lambda_\zeta}
     \frac{
           e^{-\pi\beta/2} e^{-ix} (2 x)^{-i\beta} 
}{\Gamma(\lambda_\zeta-i\beta)}
     + \text{c.~c.}
\right] (1+O(x^{-1})).
\ee

It is well known, that the necessary condition for the existence of the 
eigenvalues of the self-adjoint operator on an infinite applicable domain 
(the singular Sturm-Liouville problem) is the vanishing of the eigenfunctions 
both at the origin and at infinity.
The first condition is already satisfied by the choice of the solution 
(\ref{eq39}).
According to Eq.\ (\ref{eq40}), the second one will be satisfied in the 
gamma-function $\Gamma(\lambda_\zeta-i\beta)$ poles
\be                               \label{eq41}
\lambda_\zeta-i\beta = -n,
\ee
where $n$ is a non-negative integer.
This expression gives the discrete spectrum of eigenvalues of the energy 
operator.
The expression conjugated to Eq.\ (\ref{eq41}) leads to just the 
same spectrum:
\be                           \label{eq42}
M_{+ n}=M_- +
m \frac{
        2 m M_- (e_+^2- e_-^2 -m^2) \pm ({\cal 
N}+\zeta\alpha) \sqrt{s^2-4 m^2 M_-^2}
       }{s^2},
\ee
where ${\cal N}=2 n+1=1,~3,~5,...$, and
\be                           \label{eq43}
s^2={\cal N}^2 + 2\zeta\alpha{\cal N} +1 + 4m^2 e_-^2.
\ee

Here the radical signs ``$\pm$'' correspond to upper and 
lower energy continua and they are not directly connected neither with 
$\zeta$ 
and $\lambda_\zeta$ nor with parameters of the spacetimes outside and inside
the shell.
For the ground state $(n=0)$ we obtain
\be                           \label{eq44}
M_{+ 0}=M_- +
  \frac{m}{2}
  \frac{
        m M_- (e_+^2- e_-^2 -m^2) \pm
        2\lambda_\zeta
        \sqrt{
              \lambda_\zeta + m^2 (e_-^2 -M_-^2)
             }
       }{\lambda_\zeta + m^2 e_-^2}.
\ee

From expressions (\ref{eq33}) and (\ref{eq36}) it follows that the 
necessary conditions of the bound state existence are the inequalities
\ba
&&m^2-(\Delta M)^2 > 0,                                      \label{eq45}\\ 
&&1-m^4 + 2 m^2 
(e_+^2 +e_-^2) -(e_+^2 -e_-^2)^2 \geq 0.                     \label{eq46}
\ea

The first is the condition of energy ellipticity, which is usual for 
the existence of bound states in quantum mechanics.
The second inequality determines the extremal values of $e_\pm$ and $m$.
If it is not satisfied, a quantum instability arises \cite{bd}.
For a neutral shell the condition (\ref{eq46}) is reduced to the restriction
for the rest mass  $m \leq m_{\text{Planck}}$ \cite{hkk}.

Now we consider the special cases of quantum shells.

(i) Hollow shell ($M_-=0,~M_+=M,~e_-=0,~e_+=e$).
Equations (\ref{eq42}) and (\ref{eq44}) give, respectively,
\ba
&&M_n=\pm m
      \left[
             1+\frac{m^4}{({\cal N} +\zeta \tilde \alpha)^2} 
      \right]^{-1/2},                                         \label{eq47}\\
&&M_0=\pm  m  \sqrt{\tilde\lambda_\zeta},                     \label{eq48}
\ea
where
\be \label{eq49}
\tilde \lambda_\zeta =\frac{1+\zeta\tilde\alpha}{2},~
\tilde\alpha =\sqrt{1-(m^2 -e^2)^2}.
\ee
These spectra are the generalization of the results obtained for the neutral 
hollow shell \cite{hkk}.

(ii) Atomlike shell ($e_+=0,~e_-=e$).
One has
\be                           \label{eq50}
M_{+ n}=M_- +
m \frac{
        - 2 m M_- (e^2 + m^2) \pm
        ({\cal N}+\zeta\tilde\alpha) \sqrt{\tilde s^2-4 m^2 
M_-^2}
       }{\tilde s^2},
\ee
\be                           \label{eq51}
M_{+ 0}=M_- +
  \frac{m}{2}
  \frac{
        -m M_- (e^2+ m^2) \pm
        2\tilde\lambda_\zeta
        \sqrt{
              \tilde\lambda_\zeta + m^2 (e^2 -M_-^2)
             }
       }{\tilde\lambda_\zeta + m^2 e^2},
\ee
where $\tilde s^2 = s^2|_{\alpha=\tilde\alpha,~e_-=e}$ (\ref{eq43}).

In conclusion it should be noted that if in the case of the hollow shell 
(see Eqs.\ (\ref{eq47}) and (\ref{eq49})) we could reject the states with an 
energy from the lower continuum because of its negativity, then in the cases 
(\ref{eq50}) and (\ref{eq51}) it is impossible to do it, especially at 
large $M_-$.

\section{MONOPOLE CHARGED SHELLS} 
\label{sec-msh}

In this section we will consider shells made of the dust of
monopoles described by Eqs.\ (\ref{eq1})-(\ref{eq3}).
Special interest will be paid to their comparison with electrically 
charged shells.

Taking into account the remark after Eq.\ (\ref{eq12}), we can write 
expressions (\ref{eq7}) and (\ref{eq10}) in the form 
\be                           \label{eq52}
ds_\pm^2=-
     \left(
           1- \eta_\pm^2 - \frac{2 M_\pm}{r}
     \right) dt_\pm^2 +
     \left(
           1- \eta_\pm^2 - \frac{2 M_\pm}{r}
     \right)^{-1} dr^2 +
     r^2 d\Omega^2,
\ee
\be                           \label{eq53}
\epsilon_+ \sqrt{\dot R^2 + 1 - \eta_+^2- \frac{2 M_+}{R}  } 
-
\epsilon_- \sqrt{\dot R^2 + 1 - \eta_-^2- \frac{2 M_-}{R} } 
=  -\frac{m}{R}.
\ee
The special cases of these equations will be studied below.

\subsection{Hollow shell} 
\label{sec-msh-h}

Supposing the spacetime inside the shell to be flat 
($M_-=0,~\eta_-=0$), $M_+=M$, and $\eta_+=\eta$, we will obtain from 
Eq.\ (\ref{eq53})
\be                           \label{eq54}
\epsilon_+ \sqrt{\dot R^2 + 1 - \eta^2- \frac{2 M}{R}  } -
\epsilon_- \sqrt{\dot R^2 + 1  } =
  -\frac{m}{R}.
\ee

Analysis of this expression gives us the classification of shells, 
presented in Table \ref{tab-2}.
Unlike the hollow electric case, this classification does not depend on 
a radius (compare with Table \ref{tab-1}) and is identical to that of 
atomlike electric shells at $M_+>M_-$.

We must keep in mind that if one imposes the condition $m>0$, 
then from Table \ref{tab-2} it follows that $\epsilon_- = 1$
By means of Eq.\ (\ref{eq54}) we can prove the following proposition
in the same way as was done in the previous section.

{\it Proposition 2.}\ \
Let us have a hollow spherically symmetric dust shell with the rest 
mass $m > 0$ and monopole charge $\eta$.
Then\\
(i) the presence of the monopole charge decreases the total mass $M$ of 
shells of BH and WH types
\be \label{eq55}
M = m \sqrt{1+\dot R^2} - \frac{m^2+ \eta^2 R^2}{2 R},
\ee
(ii) the presence of the charge $\eta$ decreases the mass of black holes 
and wormholes with the throat radius equal to that of the horizon
$R_H=2 M/(1-\eta^2)$, formed by such shells,
\be \label{eq56}
M_H=\frac{m}{2}
    \left(
          1 -\eta^2
    \right).
\ee

Further, we can also consider the turning point theorem.

{\it Theorem 2.}\ \ 
Let the conditions of proposition 2 be valid.
Besides, we assume the positivity of $M$ and define
\be \label{eq57}
R_s^{(\pm)}=\frac{ m - M \pm \sqrt{(m-M)^2-m^2 \eta^2}
         }{\eta^2}.
\ee

Then for the occurrence of turning points the total mass of the moving 
monopole shell must satisfy the condition of ellipticity,
\be \label{eq58}
M < m (1-|\eta|) < m,
\ee
and 
(a) if $\epsilon_+ = \epsilon_- = 1$, then the shell has only one turning 
point $R=R^{(+)}_s$ under the total mass $M \leq M_H$ ($M_H < m/2$);
(b) if $\epsilon_+ = -\epsilon_- = -1$, then the shell can have two turning 
points, namely, $R=R^{(+)}_s$ under $M_H \leq M < m$ and
$R=R^{(-)}_s$ under $M < m$.

{\it Proof:}\ \
Under the condition $\left. \dot R \right|_{R=R_s} = 0$ from 
Eq.\ (\ref{eq55}) we obtain
\[
R_s^{(\pm)}=\frac{ - (M-m) \pm \sqrt{(M-m)^2-m^2 \eta^2}
         }{\eta^2}.
\]
The condition of radius reality yields
\[
M - m \in (-\infty,-|\eta m|] \cup [|\eta m|,+\infty).
\]
From these restrictions and Eq.\ (\ref{eq55}) it can readily be seen that
the inequality (\ref{eq58}) is valid for any $R<\infty$.

Notice also that Eq.\ (\ref{eq55}) was obtained by the 
squaring of Eq.\ (\ref{eq54}), and
therefore we must substitute $R^{(\pm)}_s$ in Eq.\ (\ref{eq54}) for checking.
Indeed, it gives us the additional constraint
\[
\text{sgn} \left( R^{(\pm)}_s -m \right) = \epsilon_+,
\]
from which it follows that
\ba
&&\epsilon_+ = 1,~~~R^{(\pm)}_s \geq m     \label{eq59}\\
&&\epsilon_+ = -1,~~R^{(\pm)}_s \leq m.     \label{eq60}
\ea
Further, one can see that the inequality
\[
R^{(-)}_s < m
\]
takes place for any $M<m$ and $\eta^2 < 1$.

Now we consider the equation
\[
R^{(+)}_s = m,
\]
the solution of which, taking into account Eq.\ (\ref{eq56}), is
\[
M = M_H.
\]
Using this solution and inequalities (\ref{eq59}) and (\ref{eq60}), we
obtain the sought-after conditions (a) and (b). Q.E.D.

\subsection{Atomlike shell. Mass amplification} 
\label{sec-msh-a}

Now we consider the case when a shell surrounds a body with an opposite
monopole charge.
Then the metrics of the spacetimes outside and inside the shell
will be described by expression (\ref{eq52}), in which 
$\eta_-=\eta,~\eta_+=0$.
From the standpoint of an observer outside the shell such a system is neutral.
Such a coupling of a charge in the neutral construction could be one more 
reason for the absence of observable monopoles.
However, keeping within the framework of the model we are not able to answer 
the ``simple'' question as to why the mechanism of the charge coupling does not 
lead to the disappearance of free electric charges.

So, according to Eq.\ (\ref{eq53}), the equations of motion read
\be                           \label{eq61}
\epsilon_+ \sqrt{\dot R^2 + 1 - \frac{2 M_+}{R}  } -
\epsilon_- \sqrt{\dot R^2 + 1 - \eta^2- \frac{2 M_-}{R} } =
  -\frac{m}{R}.
\ee
Let us classify the shells.
In the same way as was done for electrical shells, we can 
distinguish two cases.

(i) $M_+ \leq M_-$.
The classification does not depend on the radius (see Table \ref{tab-3}).                         

(ii) $M_+ > M_-$.
The classification is presented in Table \ref{tab-4}, where 
$R_{\text{extr}}=2 \Delta M/\eta^2$.
Note that, according to Ref. \cite{bv}, $\eta^2 \ll 1$ and the masses of
monopoles are predicted to be sufficiently large in comparison with 
the masses of
microscopic particles; therefore $R_{\text{extr}}$ always is large unlike its 
electric analogue.
If in the previous case (refer to Table \ref{tab-1}) we can 
conditionally separate the shells into those of microscopic and ordinary 
scales, then here the
division into ordinary and cosmological shells is rather suitable.

Let us find the mass of thin-shell black holes and wormholes with the horizon
radius.
Supposing $R=2 M_+$ (i.e., an observer is assumed to be outside the shell), 
from Eq.\ (\ref{eq61}) we obtain a square equation for their mass:
\be \label{eq62}
(\eta^2 - 1) M_+^2 + M_+ M_- + \frac{m^2}{4} =0,
\ee
which has solutions at $\eta^2 < 1$.
Only one of them is positive
\be \label{eq63}
M_+ = \frac{1}{2 (1-\eta^2)}
      \left(
            M_- + \sqrt{M_-^2 + m^2 (1-\eta^2)}
      \right) .
\ee

It is easy to see that when increasing $\eta$ up to $1$, the observable 
mass increases infinitely.
However, according to the aforesaid, there is no physical meaning beyond it.
Actually, it is necessary to consider this expression at $\eta^2 \ll 1$, 
i.e.,
\be \label{eq64}
M_+ =
      M_{\text{neutral}}
      \left(
            1 +
            \frac{M_{\text{neutral}}
                 }
                 {\sqrt{M_-^2 + m^2}} \eta^2 
\right) + O(\eta^4),
\ee
where $M_{\text{neutral}} = \left(M_- + \sqrt{M_-^2 + m^2}\right)/2$ 
is the observable mass of the system at $\eta =0$.

From this expression it is obvious that the effect of the amplification 
of the internal mass by shell monopole charge takes place.
This effect is opposite to that of screening studied in the electric case
(refer to Eqs.\ (\ref{eq26})-(\ref{eq28}) and commentary after them).

\subsection{Quantum shells} 
\label{sec-msh-q}

Similarly to Sec. \ref{sec-esh-q}, one considers the quantum properties for
the most general case (\ref{eq52}); after that the hollow and atomlike 
shells will be considered as special cases.
Everywhere, unless otherwise qualified, the definitions of 
Sec. \ref{sec-esh-q} are used.

So it is possible to rewrite Eq.\ (\ref{eq53}) in the form of 
the energy-momentum conservation law for a point particle 
(cf. Eq.\ (\ref{eq29})):
\be         \label{eq65}
\left(
      \Delta M + \frac{(\eta_+^2 -\eta_-^2)}{2} R
\right)^2 -\Pi^2 =
m^2
\left(
1 - \frac{(\eta_+^2 +\eta_-^2)}{2} - \frac{M_+ + M_-}{R} -
\frac{m^2}{4 R^2}
\right).
\ee

The equation for the wave function of the shell is obtained, as was 
done for Eq.\ (\ref{eq30}):
\be   \label{eq66}
R^2 \Psi^{\prime\prime} +
\left[
       k^2 R^4
       + 2 k \Delta M R^3 +
       \chi^2 R^2 + a R + b
\right]\Psi =0,
\ee
where
\ba
&&\chi=\sqrt{
             (\Delta M)^2-m^2
             \left(
                   1 - \frac{\eta_+^2 +\eta_-^2}{2}
             \right)
            },                                           \nn\\
&&a=m^2 (M_+ +M_-),                                      \label{eq67}\\
&&4 b= m^4,~k = \frac{\eta_+^2 -\eta_-^2}{2}.                    \nn
\ea
Let us consider the asymptotics.\\
(a) $R\rightarrow\infty$. 
We have the equation
\be \label{eq68}
\Psi^{\prime\prime} + (k R)^2\Psi =0,
\ee
the solution of which is expressed in the Bessel functions 
\[
\Psi =  \sqrt{R}
        \left[
              c_1 \BesselJ{1/4}{| k | R^2} + c_2
\BesselY{1/4}{| k | R^2}
        \right]
\]
\be \label{eq69}
\rightarrow
\sqrt{
      \frac{2}{\pi | k | R}
     }
        \left[
           c_1 \cos{\left(
                          | k | R^2 -
                          \frac{3 \pi}{8}
                    \right)
                   } +
           c_2 \sin{\left(
                          | k | R^2 -
                          \frac{3 \pi}{8}
                    \right)
                   }
        \right].
\ee
(b) $R\rightarrow 0$.
We have
\ba
&&R^2 \Psi^{\prime\prime} + b\Psi =0,
\label{eq70}\\
&&\Psi = c_+ R^{\lambda_{+1}} + c_- R^{\lambda_{-1}},
\label{eq71}
\ea
where $c_{1,2,+,-}$ are integration constants,
\ba
&&\lambda_\zeta =\frac{1+\zeta\alpha}{2},~\zeta=\pm 1, \nn\\
&&\alpha = \sqrt{1-m^4}.                                \label{eq72}
\ea

From the last expression it is evident that bound states can exist only at
$m \leq 1$, as took place for the hollow neutral 
shell \cite{hkk}.\footnote{See also the comments after inequality (\ref{eq46}).}
Thus we obtain a restriction for the mass of quantum thin-shell monopoles.

Generally speaking, the exact solution (\ref{eq66}) cannot be found in 
known functions.
Fortunately, we do not need it.
We will consider the approximation for small monopole charges, which
corresponds to the physical picture \cite{bv,dny,kms}.
Neglecting the terms with $k^2$ and $k \Delta M$, we obtain an
equation like Eq. (\ref{eq30}), where $\chi,~a$, and $b$ are defined by 
Eqs.\ (\ref{eq67}).
Furthermore, we can rewrite this equation as Eq.\ (\ref{eq37}), where
\be \label{eq73}
\beta=\frac{m^2 (M_+ +M_-)}{2 \chi},
\ee
and obtain the solutions (\ref{eq39}).
By analogy with Eq.\ (\ref{eq42}) we obtain the discrete spectrum of the 
bound states of our generalized monopole shell:
\be                           \label{eq74}
M_{+ n}=
  \frac{
        M_- (s^2- 1) \pm s
        \sqrt{m^2 ( s^2+1) (1-\frac{\eta_+^2 +\eta_-^2}{2}) 
- 4 M_-^2}
       }{s^2 +1},
\ee
where
\be                           \label{eq75}
s=\frac{{\cal N} + \zeta\alpha}{m^2}.
\ee
The remarks about the radical signs ``$\pm$''  are given after 
expressions (\ref{eq43}) and (\ref{eq51}).\\ 
Let us study the special cases.

(i) Hollow shell ($M_-=0,~M_+=M,~\eta_- =0,~\eta_+=\eta$).
Equation (\ref{eq74}) gives the spectrum
\be          \label{eq76}
M_n=\pm m \sqrt{1-\frac{\eta^2}{2}}
      \left[
             1+\frac{m^4}{({\cal N} +\zeta \alpha)^2} 
\right]^{-1/2},
\ee
which is different from that of the neutral shell by the scale factor 
$\sqrt{1-\eta^2/2}$.
For the ground state we obtain
\be     \label{eq77}
M_0=\pm m \sqrt{\lambda_\zeta (1-\eta^2/2)}.
\ee

(ii) Atomlike shell ($\eta_+=0,~\eta_-=\eta$).
One has
\be                           \label{eq78}
M_{+ n}=
  \frac{
        M_- (s^2- 1) \pm s
        \sqrt{m^2 ( s^2+1) (1-\frac{\eta^2}{2}) - 4 M_-^2}
       }{s^2 +1},
\ee
\be                           \label{eq79}
M_{+ 0}=\zeta\alpha M_- \pm m
        \left[
              \lambda_\zeta
              \left( 1-\frac{\eta^2}{2} \right)
              -m^2 M_-^2 
        \right]^{1/2}.
\ee

Analyzing the results of this section, it is possible to make a 
conclusion that the properties of the shells related to the Barriola-Vilenkin 
metric are near to those of the Schwarzchild shells.
It does not seem to be extraordinary, because these metrics are very cognate;
e.g., the Penrose diagram for the metric (\ref{eq3}) is identical to the 
Schwarzchild one with a redefinition of the black hole surface 
gravity \cite{dny}.

\section{conclusion}
\label{sec-con}

Let us summarize the main results of the investigation and comparison of the 
electrically and monopole charged shells.

For the hollow shells it was shown that 
(1) electric charge increases the total mass-energy of moving shells
as well the mass of the thin-shell black holes and traversable wormholes 
formed by them;
(2) monopole charge decreases the total mass-energy of moving shells
as well the mass of the thin-shell black holes and wormholes.

Also we proved the turning point theorems, which are necessary, e.g., for a
correct determination of forbidden and permitted motion regions that can
be useful for WKB quantization, shell decay or nucleation analysis, etc.
It should be noted that a key common feature of these theorems is 
the requirement of ellipticity of the observable mass-energy $M$.
In addition, the theorems give us more detailed properties of the 
studied shells.

As for the atomlike configurations, special interest had been paid to 
the nonlinear dependence of the observable mass of a whole system on the 
masses and charges of constituent bodies.
Such a dependence leads to some effects.
There are the amplification (monopole case) and screening (electrical case) 
of an internal mass $M_-$ by a shell charge.
We considered these effects for two significance levels, namely, for 
moving shells in the general case and shells with horizon radii.

For every shell studied  we have constructed a classification, from which 
the division of shells into those of microscopic, ordinary, or 
astronomical scales resulted in a natural way. 

Finally, we have considered quantum shells.
Exact solutions of the wave equations and mass spectra of bound states have 
been found in this paper for the several physically interesting cases.
In particular, it is shown that the rest mass of monopole quantum shells
cannot be in excess of the Planckian one, as takes place for the 
neutral shell.

\begin{acknowledgments}
I am profoundly grateful to Professor V.D. Gladush for helpful discussions.
Also I would like to thank R.A. Konoplya for a reading of the manuscript.
\end{acknowledgments}

\begin{table}
\caption{
This classification is suitable for both hollow and atomlike 
(at $M_+ < M_-$) electrically charged shells.
The signs ``BH'' and ``WH'' denote the existence of shells of 
black hole and wormhole types, respectively; the sign ``$\star$'' denotes 
the impossibility of a junction.}
\begin{tabular}{ccccc}
 &\multicolumn{2}{c}{$\epsilon_+ =\epsilon_-$}
 &\multicolumn{2}{c}{$\epsilon_+ =-\epsilon_-$}\\
 &${\epsilon_+=1\choose{\epsilon_-=1}}$ 
  &${\epsilon_+=-1\choose{\epsilon_-=-1}}$ &${\epsilon_+=1\choose{\epsilon_-
=-1}}$ &${\epsilon_+=-1\choose{\epsilon_-=1}}$\\
 \tableline
 $R < R_{\text{extr}}$ &$\star$&BH &$\star$ &WH \\
 $R = R_{\text{extr}}$ &$\star$&$\star$&$\star$ &WH \\
 $R > R_{\text{extr}}$ &BH     &$\star$&$\star$&WH \\
 \end{tabular}
 \label{tab-1}
 \end{table}

\begin{table}
\caption{
This classification is suitable for both the atomlike electrically charged 
shell (at $M_+ \geq M_-$) and hollow monopole charged shell.}
\begin{tabular}{ccccc}
 &\multicolumn{2}{c}{$\epsilon_+ =\epsilon_-$}
 &\multicolumn{2}{c}{$\epsilon_+ =-\epsilon_-$}\\
 &${\epsilon_+=1\choose{\epsilon_-=1}}$ 
  &${\epsilon_+=-1\choose{\epsilon_-=-1}}$ 
 &${\epsilon_+=1\choose{\epsilon_-=-1}}$ 
  &${\epsilon_+=-1\choose{\epsilon_-=1}}$\\
 \tableline
 $R > 0$ &BH     &$\star$&$\star$&WH \\
\end{tabular}
\label{tab-2}
\end{table}

\begin{table}
\caption{
The classification of atomlike monopole charged shells at
$M_+ \leq M_-$.}
\begin{tabular}{ccccc}
 &\multicolumn{2}{c}{$\epsilon_+ =\epsilon_-$}
 &\multicolumn{2}{c}{$\epsilon_+ =-\epsilon_-$}\\
 &${\epsilon_+=1\choose{\epsilon_-=1}}$ &${\epsilon_+=-
1\choose{\epsilon_-=-1}}$ &${\epsilon_+=1\choose{\epsilon_-
=-1}}$ &${\epsilon_+=-1\choose{\epsilon_-=1}}$\\
 \tableline
 $R > 0$ &$\star$&BH&$\star$&WH \\
\end{tabular}
\label{tab-3}                                   
\end{table}                                     
                                                
\begin{table}                                   
\caption{
The classification of atomlike monopole charged shells at $M_+ > M_-$.}
\begin{tabular}{ccccc}
 &\multicolumn{2}{c}{$\epsilon_+ =\epsilon_-$}
 &\multicolumn{2}{c}{$\epsilon_+ =-\epsilon_-$}\\
 &${\epsilon_+=1\choose{\epsilon_-=1}}$ &${\epsilon_+=-
1\choose{\epsilon_-=-1}}$ &${\epsilon_+=1\choose{\epsilon_-
=-1}}$ &${\epsilon_+=-1\choose{\epsilon_-=1}}$\\
 \tableline
 $R < R_{\text{extr}}$ &BH&$\star$&$\star$ &WH \\
 $R = R_{\text{extr}}$ &$\star$&$\star$&$\star$ &WH \\
 $R > R_{\text{extr}}$ &$\star$&BH&$\star$&WH \\
 \end{tabular}
 \label{tab-4}
 \end{table}

\def\CMPh{Commun. Math. Phys.}
\def\JPh{J. Phys.}
\def\CJP{Czech. J. Phys.}
\def\LMPh {Lett. Math. Phys.}
\def\NPh  {Nucl. Phys.}
\def\PhE  {Phys.Essays}
\def\PhL  {Phys. Lett.}
\def\PhR  {Phys. Rev.}
\def\PhRL {Phys. Rev. Lett.}
\def\PhRp {Phys. Rep.}
\def\NCim {Nuovo Cimento}
\def\NuPB {Nucl. Phys.}
\def\GRG {Gen. Relativ. Gravit.}
\def\CQG {Class. Quantum Grav.}
\def\prp {report}
\def\Prp {Report}

\def\jn#1#2#3#4#5{{#1}{#2} {\bf #3}, {#4} {(#5)}}

\def\boo#1#2#3#4#5{{\it #1} ({#2}, {#3}, {#4}){#5}}

\def\prpr#1#2#3#4#5{{``#1,''} {#2 }{#3}{#4}, {#5} (unpublished)}


\begin{references}
\bibitem{eq17proof}
Expression (\ref{eq17}) can easily be found if one substitutes the external 
spacetime $\Sigma^+$ horizon radii 
$R_{H (1,2)} = M + \text{sgn}(\sqrt{M^2 - e^2}) \sqrt{M^2 - e^2}$ in 
Eq.\ (\ref{eq15}).
Herewith, after rejection of superfluous roots we find also that the equality
$\text{sgn}(\sqrt{M^2 - e^2}) = \epsilon_-$ takes place.
Further, all masses are assumed to be positive; therefore $\epsilon_- = 1$
and we must choose only the external horizon 
$M + \sqrt{M^2 - e^2}$ as the radius of the collapsed shell.
\end{references}
\end{document}